\documentclass[preprint,showpacs,preprintnumbers,amsmath,amssymb]{revtex4-1}

\usepackage{graphicx}
\usepackage{dcolumn}\usepackage{bm}
\usepackage{amssymb}

\newcommand{\sbentcore}{bent-core }
\newcommand{\density}{ \rho }
\newcommand{\opdfrho}{ p }
\providecommand{\n}{\nonumber \\ }
\providecommand{\orderparam}[3]{\overline{\Delta^{(#1)}_{#2,#3}}}
\providecommand{\dg}[1]{#1^{\circ}}
\providecommand{\lnbra}{\left(}
\providecommand{\rnbra}{\right)}

\begin{document}

\title{Stability of Biaxial Nematic Phase in Model Bent-Core Systems}

\author{Piotr Grzybowski}
\email[e-mail address:]{merlin@th.if.uj.edu.pl}
\affiliation{Marian Smoluchowski Institute of Physics, Department
of Statistical Physics and  Mark Kac Center for Complex
Systems Research, Jagellonian University, Reymonta 4,
Krak\'ow, Poland}
\author{Lech Longa}
\email[e-mail address:]{lech.longa@uj.edu.pl}
\affiliation{Marian Smoluchowski Institute of Physics, Department
of Statistical Physics and  Mark Kac Center for Complex
Systems Research, Jagellonian University, Reymonta 4,
Krak\'ow, Poland}

 \date{\today}

\begin{abstract}
 We study a class of models for \sbentcore molecules
using low density version of Local Density Functional Theory.
Arms of  the molecules are modeled  using two- and three Gay-Berne
(GB) interacting units of uniaxial and biaxial symmetry. Dipole-dipole interactions
are taken into account by placing a dipole moment along the ${\mathcal{C}}_2$
symmetry axis of the  molecule. The main aim of the study is to identify molecular
factors that can help stabilizing the biaxial nematic phase.
The phase diagrams involving isotropic ($I$), uniaxial ($N_U$)
and biaxial  ($N_B$) nematic phases are determined
at given density and dipole strength  as function of bent angle.
For molecules composed of two uniaxial arms a direct $I-N_B$
phase transition is found at  a single Landau point,
which moves towards lower bent angles with increasing dipole magnitude.
For the three-segment model strengthening of the dipole-dipole interaction
results in appearance of a line of Landau points.  There exists an optimal
dipole strength for which this line covers the maximal  range of opening angles.
Interestingly,  the inclusion of biaxial GB ellipsoids as building blocks  reveals
the direct $I-N_B$ transitions line even  in a non-polar, two-arms model.
The line is shifted towards  higher opening  angles as compared to the
uniaxial case.
\end{abstract}

 \maketitle

 Thermotropic biaxial nematic phase, embodying the "holy grail" \cite{luckhurst}
of the liquid crystals physics, recently has engendered much scientific interest.
In 2004 a long-awaited discovery was announced \cite{madsen,acharya,severing}
and the thermotropic biaxial nematics were claimed to be found  at last,
34 years after its first theoretical prediction by Freiser in 1970 \cite{freiser}. The discovery was
made in systems of the \sbentcore mesogens, also often called  banana- or boomerang-shaped. They possessed a rigid bent-core with apex angle of $\sim\hspace{-1.5mm}140^{\circ}$, although  recently the biaxial phase has also been reported
for  bent-core systems with apex angle of $90^{\circ}$  \cite{mlehmann}.
It should be noted that first experimental speculations of possible
biaxiality in \sbentcore systems were made earlier \cite{earlybentcorebx}.

 From theoretical point of view it had been known that one can expect
a biaxial phase in \sbentcore systems. Bifurcation analysis for hard
boomerang model, with molecules composed of two hard spherocylinders
joined at their ends, was
carried out by Teixeira, Masters and Mulder \cite{teixeira}
predicting phase diagrams with a direct $I-N_B$ phase transition at an
isolated Landau point for apex angle of $107^{\circ}$.
 Subsequent mean-field analysis of interacting quadrupoles
 by Luckhurst \cite{luckhursttsf} have
predicted  that only a
bond angle within a few degrees of the tetrahedral
angle ($109.47^{\circ}$), where Landau point was observed,
would allow the biaxial nematic to appear above the
freezing point of a real uniaxial nematic.
That raised a new question, since the experiments
reported the angle to be near $140^{\circ}$. Among others that issue
has been addressed in various simulation approaches, including
Metropolis Monte Carlo study of a Lebwohl-Lasher lattice model
\cite{mcasymmetric}. Introduced asymmetry of molecule arms
in the aforementioned paper
shifted Landau point towards lower angles. The same model was used
to investigate the influence of molecule flexibility
\cite{mcflexible,longaFlexible}, and recently of dipole-dipole interactions \cite{mcbatesdipoles}
on phase diagram.

As it has appeared quite recently, the biaxial phase proves to be more challenging experimentally than earlier thought \cite{le}. It seems therefore important to look into molecular models that can help in a proper understanding the biaxiality issue. In the present paper we construct a class of models of bent-core molecules with the aid of including a broad range of molecular factors that can appear relevant
in stabilizing biaxial phase. We build up  the  molecule from parts interacting through  Gay-Berne  (GB) potential \cite{gay} and
go beyond lattice models to account, at least partly, for excluded volume effects. We will keep the arm's length close to experimental values.
The simulations for similar molecules, but with relatively short arms,
have not shown any trace of the biaxial nematic phase \cite{johnston,johnston2,memmer,orlandi}.
Apart from isotropic and uniaxial nematic phase, smectic phases were observed, which is in agreement with predictions of hard spherocylinder dimer model
presented in \cite{mchardspherocylinder}.

 The atomistic simulations \cite{atomistic} of the molecular systems studied experimentally
\cite{madsen,acharya} have confirmed the existence of a  weak
biaxiality in that system. Similarly,  for a  system of  GB bent-core
molecules with  long arms the biaxial nematic phase was
observed on cooling \cite{Jozefowicz}.

\begin{figure}
\includegraphics{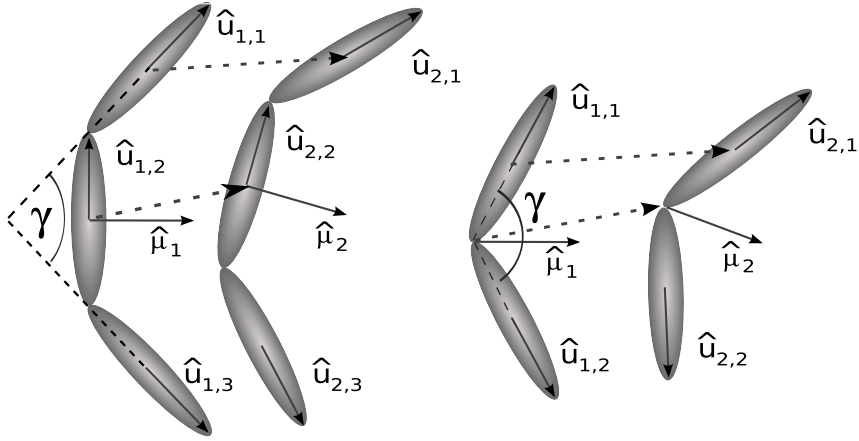}%
\caption{
 \label{bananasconstr}
  The construction of pair potential out of two and three GB interacting parts
  firmly attached to molecule.
  Each part of one molecule interacts with every part from other molecule via $V_{GB}$.
  Dipole-Dipole  interaction, $V_{DD}$, is also added.
 }
\end{figure}

Our attempt in the present work is to  employ the  GB interaction\cite{gay,bfz}
to model interacting parts of the molecules.
Then, the models are studied using low density approximation to
the Local Density Functional theory \cite{evans,hansen}.
 By means of bifurcation analysis \cite{mulder,longa},
we determine  bifurcation phase diagrams for bent-cores
constructed out of two and three GB parts, as shown
in Fig.~\ref{bananasconstr}.
Finally, we introduce   the dipole-dipole interactions  with the dipoles
taken parallel to molecular ${{\mathcal{C}}_2}$ symmetry axis and investigated the influence of
the dipole strength on stability of  $N_B$.

\begin{figure}
\includegraphics{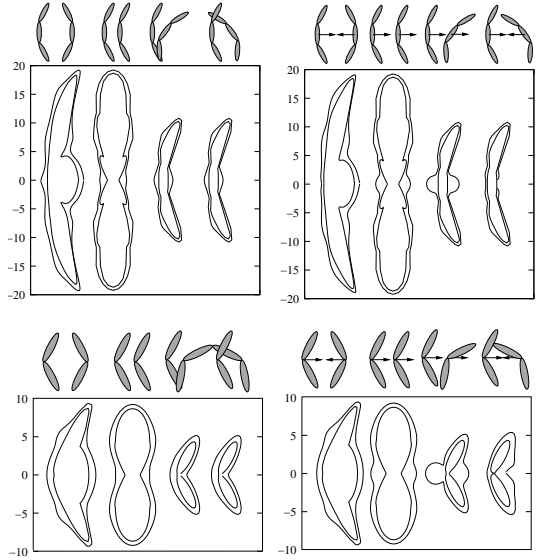}%
 \caption{
  \label{bananasprofile}
  Exemplary  equipotential surfaces for model bent-core molecules composed of 2 ( bottom)
  and 3 (top) uniaxial parts for arm's elongation of $5:1$ and  for opening angle
  $\gamma=126^{\circ}$.
  The case with the dipole-dipole interaction included ($\mu=2.0$) is shown to the right.
  Surfaces are shown for
  the total pair potential equal to $0$ and $-0.2$.
 }
\end{figure}

We employ reduced units  by  setting $\sigma_{0}=1$ and  $\epsilon_{0}=1$ for the
GB potential parameters \cite{gay}. The reduced distance $r^{*}$ and the
reduced temperature $t $ are then given by   $r^{*} = r/\sigma_{0}$ and
$t = k_{B}T/\epsilon_{0}$, respectively.
We also set $\nu=1$, $\mu=2$  \cite{gay} and choose the ratio of
length to breath of 5:1 for the uniaxial arms.

The  biaxial arms of  bent-core molecules  are modeled with the help of soft GB  ellipsoids
as proposed by Berardi-Fava-Zannoni   \cite{bfz},
where for the axes we take $\lnbra \sigma_{x},\sigma_{y},\sigma_{z} \rnbra=\lnbra 1.2,0.514,3.4\rnbra$
 and for the potential parameters $\lnbra \epsilon_{x},\epsilon_{y},\epsilon_{z}\rnbra=\lnbra 1.0,1.4,0.2\rnbra$; in addition $\mu=1$  and $\nu=3$ (for definitions of the parameters please refer to \cite{bfz}).
 The ellipsoids are oriented so that the shortest axis is perpendicular to the molecular symmetry plane containing molecular parts and the longest one is lying in that plane. As mentioned in \cite{bfz} those parameters and orientation of the ellipsoids
make the attractive forces strongest in the
face-to-face configuration (that is in the direction of shorter axis).
Clearly, that model exhibits molecular biaxiality in the limit
of $\gamma=\dg{180}$ and $\gamma=0.0$; there should also exist an angle separating rod-like and disk-like molecular shapes.

 The resulting GB potential  $V_{GB}$  for a pair of bent-core molecules is then a
sum of four  terms for the two-part case  and of nine terms for the three-part case.  The  dipole-dipole part of interaction is of standard form:
\begin{eqnarray}
 V_{DD}\left(\boldsymbol{\mu}_{1},\boldsymbol{\mu}_{2},\bf{r}\right) =
  \frac{ \boldsymbol{\mu}_{1}\cdot\boldsymbol{\mu}_{2}
        - 3 \left( \boldsymbol{\mu}_{1}\cdot\bf{r} \right)
        \left( \boldsymbol{\mu}_{2}\cdot\bf{r} \right) }
        { r^{3} } \nonumber \, ,
\end{eqnarray}
 where $\boldsymbol{\mu}_{i}=\mu \hat{\boldsymbol{\mu}}_{i} \, , i=1,2$;  $\hat{\boldsymbol{\mu}}_{i}$ is the unit vector and $\mu$  the magnitude of  $\boldsymbol{\mu}_{i}$. The total pair interaction, $V$, is the sum of $V_{GB}$ and $V_{DD}$.
 In Fig.~\ref{bananasprofile} shown are  the exemplary equipotential surfaces.
 Figures show the profiles for four
different relative molecular orientations, including two- and three GB part bent-core molecules
of uniaxial arms.

 Relative importance of $V_{GB}$  and $V_{DD}$  parts is measured by
 by the ratio $\left|\frac{V_{DD}}{V_{GB}+V_{DD}}\right|$. The   ground state values of this parameter for different values of $\mu$ are given  in Table~\ref{muvsdd}.
\begin{table}
 \caption{
  \label{muvsdd}
  Dipole-dipole contribution to total pair potential energy in ground state.
 }
 \begin{ruledtabular}
  \begin{tabular}{cc}
   $\mu$ & $\left|V_{DD}/\lnbra V_{GB}+V_{DD}\rnbra\right|$ \n
   $2.8$ & $0.50$ \n
   $2.2$ & $0.40$ \n
   $1.6$ & $0.25$ \n
   $1.5$ & $0.22$ \n
   $1.2$ & $0.15$
  \end{tabular}
 \end{ruledtabular}
\end{table}
We study the equilibrium properties of the systems by minimizing
the grand potential with respect to the   one-particle
distribution function $\opdfrho$ \cite{longa,evans,hansen,mulder}.
The necessary condition is given by  a self-consistent integral  equation
for stationary distribution  $\opdfrho_{S}(q)$. It reads \cite{longa}
 \begin{equation}\label{selfconsistent}
  \opdfrho_{S}(q) = Z_S^{-1}\, \exp\{\,C_{1}(q,[\opdfrho_{S}]) \},
 \end{equation}
 where  $Z_S=\int \exp\{\,C_1(q,[\opdfrho_{S}])\} \mathrm{d} q /\langle N \rangle$ is the normalization constant,
$C_1(q,[\opdfrho_{S}])$ is  the one-particle direct correlation function,
$\langle N \rangle$  is  the average number of particles in the system and $q$ represents collectively all degrees of freedom of a single molecule.

 In the present study we  (a) approximate $C_1$  using second order virial expansion,
sometimes also referred to as  low density approximation and  (b) restrict analysis to nematics,
which amounts in  setting  $\opdfrho_{S}(q)=\density P(\mathbf{\Omega})$, where
$\density= \frac{<N>}{V}\sigma^{3}_{0}$ is the  dimensionless density and
$\mathbf{\Omega}$  are the Euler angles parameterizing orientation of a molecule-fixed
frame with respect to the laboratory-fixed frame. After these limitations the Eq.~(\ref{selfconsistent})
becomes:
\begin{equation}
 \label{ldselfconsistent}
 P(\mathbf{\Omega_1}) = Z^{-1}\, \exp \left[ \density
           \int c_2\left(\mathbf{\Omega}_{1} \mathbf{\Omega_2} \right)
	   P\left( \mathbf{\Omega}_{2} \right) \mathrm{d}\mathbf{\Omega}_{2}
	   \right] \, ,
\end{equation}
 with
\begin{eqnarray}
c_2\left(\mathbf{\Omega}_{1} \mathbf{\Omega}_{2}\right) \equiv
         \int \left\{
         \exp\left[-\frac{1}{t} V(\mathbf{\Omega}^{-1}_{1} \mathbf{\Omega}_{2},\bf{r}^{*}_{12})\right]-1
         \right\}
	 \mathrm{d}^{3}\bf{r}^{*}_{12} \, , \nonumber
\end{eqnarray}
 and normalization constant:
\begin{equation}
 \nonumber
 Z = \int \exp\left[ \density \int c_2\left(\mathbf{\Omega}_{1} \mathbf{\Omega}_{2} \right)
     P\left( \mathbf{\Omega}_1 \right) \mathrm{d}\mathbf{\Omega}_{1} \right] \mathrm{d}\mathbf{\Omega}_{2}
     \, .
\end{equation}
Here  $ \mathrm{d}\mathbf{\Omega}= \mathrm{d}\alpha\,\mathrm{d}\left(\cos\left(\beta\right)\right)\mathrm{d}\gamma$
 stands for integration over Euler angles,   $\mathrm{d}^{3}{\bf{r}^{*}}= r^{*2}\mathrm{d}r^{*}\mathrm{d}\left(\cos\left(\theta\right)\right)\mathrm{d}\phi$
 and $\mathbf{\Omega}^{-1}_{1} \mathbf{\Omega}_{2}$ is the relative orientation of the molecules.

 The usual expansions of $P(\mathbf{\Omega})$ in the base of $D_{2h}$ symmetry adapted $\Delta$
functions \cite{mulder}
\begin{eqnarray}\label{OPDFKERN}
 P(\mathbf{\Omega}) & = & \sum_{L,m,n}\frac{2L+1}{8\pi^2}
 \overline{\Delta^{(L)}_{m,n}}
 \Delta^{(L)}_{m,n}(\mathbf{\Omega})
\end{eqnarray}
allows to introduce the order parameters  $\overline{\Delta^{(L)}_{m,n}}  =  \int
 \mathrm{d}\mathbf{\Omega}\,
 P(\mathbf{\Omega})\,\Delta^{(L)}_{m,n}(\mathbf{\Omega})  $ for nematics.
 The summation  runs over the allowed values of \{$L,m,n$\}, where  $L$ is a non-negative integer.
 Generally,  if $L$ is even, then $0\le m \le  L $ and $0\le n \le L$. If  $L$ is odd, then $2\le m \le L$ and
$2\le n \le L$.    If, in addition, we expand  $c_{2}$:
\begin{eqnarray}
 c_{2}(\mathbf{\Omega}_{1}\mathbf{\Omega}_{2}) & = &
 \sum_{L,m,n} c_{mn} \Delta^{(L)}_{m,n}(\mathbf{\Omega}^{-1}_{1}\mathbf{\Omega}_{2})  \,
 \nonumber
\end{eqnarray}
where
\[c_{mn}  =  \frac{2L+1}{8\pi^2} \int \mathrm{d}\mathbf{\Omega}^{-1}_{1}\mathbf{\Omega}_{2}\, c_{2}(\mathbf{\Omega}_{1}\mathbf{\Omega}_{2}) \,
 \Delta^{(L)}_{m,n}(\mathbf{\Omega}^{-1}_{1}\mathbf{\Omega_{2}})\]
 and where  $c_{mn}=c_{nm}$   due to  particle interchange symmetry,
 then  Eq.~(2) becomes reduced  to the set (in general infinite) of nonlinear equations for
 the order parameters. Using bifurcation analysis  we now seek for  a subset of nonzero
 order parameters, describing
 low-symmetry phase,  that branch off from the background high-symmetry phase.
 Generally,  in the isotropic phase all order parameters vanish.
 The uniaxial phase is characterized by nonzero  order parameters indexed by $m =0$.
 Finally, in the biaxial nematic phase all the order parameters become nonzero.

 The  bifurcation points so determined  are either spinodal points for the first-order phase transitions
or critical points for the continuous transitions. Hence, for continuous  and weakly first-order
phase transitions, as holds of  isotropic and nematic phases,
we arrive at quite accurate  estimates of the  phase diagrams in temperature-density plane.
Following the analysis as described in \cite{longa} two different bifurcation formulas can derived. The first one is the equation for the bifurcation from the isotropic phase. It reads
\begin{equation}
 \label{fromiso}
 \density = \frac{10}{c_{00}+c_{22}-\sqrt{4 c_{02}^{2} \left( c_{00} -
                   c_{22}\right)^{2}}} \, .
\end{equation}
For the uniaxial to biaxial bifurcation a more complex formula is found
\begin{eqnarray}
 \label{fromuni}
 \density & = & 35 \left[ \frac{ 2 c_{00}\left(a d c_{02}-d\right)+c_{02}\left(2 a+a b c_{22}\right)-d^{2} c_{00}+b c_{22} }{\left( a^{2}+2 b d \right) \left( c_{02}^{2} + c_{00}c_{22} \right)} \right. \n
 & - & \left. \frac{\sqrt{a^{2} c_{00}c_{22}+b d \left( c_{00}c_{22}-2 c_{02}^{2} \right)}}{\left( a^{2}+2 b d \right) \left( c_{02}^{2} + c_{00}c_{22} \right)} \right],
\end{eqnarray}
 where
 \begin{eqnarray}
  \nonumber
  a & = & 20 \orderparam{2}{0}{2}+\sqrt{15}\orderparam{4}{0}{2}, \nonumber \\
  b & = & 14+20\orderparam{2}{0}{0}+\orderparam{4}{0}{0}+\sqrt{35}\orderparam{4}{0}{4}, \nonumber \\
  d & = & 7-10\orderparam{2}{0}{0}+3\orderparam{4}{0}{0}, \nonumber
 \end{eqnarray}
and where $\orderparam{L}{m}{n}$ are determined in the uniaxial nematic phase.
 We would like to point out that since the formulas for
$\orderparam{L}{m}{n}$s depend on $\density$, the equation (\ref{fromuni})
for given $t$ becomes a self-consistent equation for density.

\label{appendix}

Please note that each $c_{mn}$ is a six dimensional integral. We performed
numerical integration to obtain temperature dependence for
each of the coefficients for given set of the molecular parameters.
The integration procedure was implemented in C.
 We have incorporated both Monte Carlo and adaptive multi-point Gauss
quadratures method in the integration procedure.
We performed Monte Carlo (MC) integration over orientations using quaternion parametrization of rotations and then calculated  the integral over
length of intermolecular vector $\bf{r_{12}}$ using adaptive Gauss quadratures.
Approximately $2$ million of MC cycles were used to calculate the integral;
the relative error was estimated to be less than $1$\%.
\begin{figure}
 \includegraphics{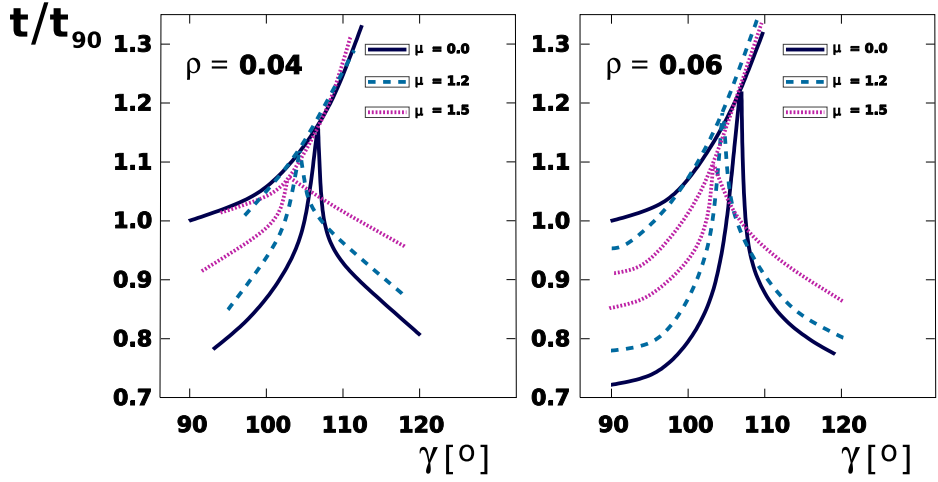}%
 \caption{
  \label{diabramb2}
   Diagrams for two-parts banana, for two densities ($\rho$)
and three values of dipole moment ($\mu$). On each plot two branches of bifurcation from
uniaxial to biaxial phase meet the upper line of bifurcations from isotropic phase.
 }
\end{figure}
\begin{figure}
 \includegraphics{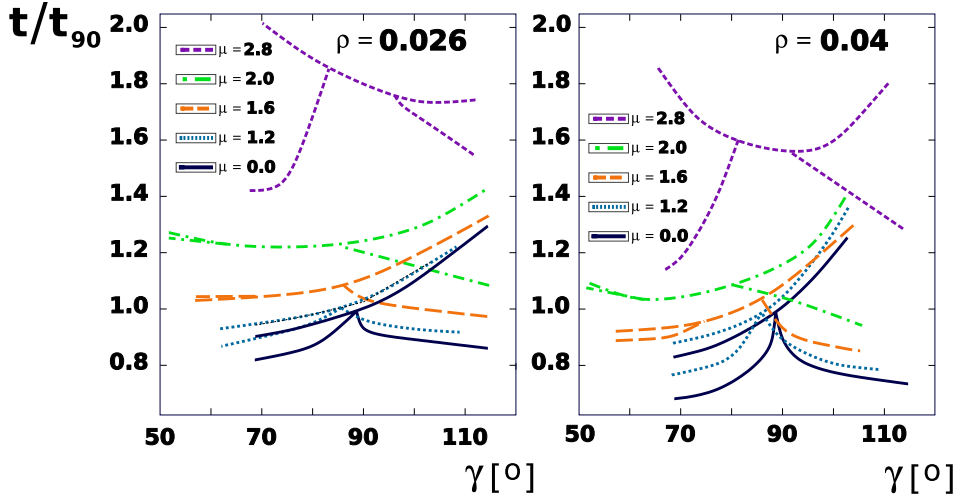}%
 \caption{
  \label{diagramb3}
   Diagrams for three-parts banana, for two densities, and five dipole moments. Each plot shows the bifurcation from isotropic phase and two lower branches of uniaxial-to-biaxial phase transition.
 }
\end{figure}
 Now we present the bifurcation diagrams, which follow from the
solutions of the bifurcation equations (\ref{fromiso}) and (\ref{fromuni}) for the uniaxial $V_{GB}$.  In Fig.~\ref{diabramb2} and \ref{diagramb3}
we have plotted $t$ divided by $t_{90}$, the temperature of bifurcation
from isotropic phase for $\gamma=90^{\circ}$ in the non-polar case.
Numerical values of $t_{90}$ are gathered in Table~\ref{t90s}.
\begin{table}
 \caption{
  \label{t90s}
  Bifurcation temperatures from isotropic phase for $\gamma=90^{\circ}$.
 }
 \begin{ruledtabular}
  \begin{tabular}{cc}
   $\density$ & $t_{90}$ \n
   \sbentcore molecule composed of two parts: & \n
   $0.04$  & $0.84$ \n
   $0.06$  & $0.96$ \n
   \sbentcore molecule composed of three parts: & \n
   $0.026$  & $1.34$ \n
   $0.04$  & $1.71$ \n
  \end{tabular}
 \end{ruledtabular}
\end{table}
 Note that the standard phase sequence \cite{teixeira} is recovered that involves the isotropic phase,
the rod-like and disc-like uniaxial nematic phase and the biaxial nematic phase.
\begin{table}
 \caption{
  \label{landauvsmu}
  Landau point versus dipole magnitude $\mu$
 }
 \begin{ruledtabular}
  \begin{tabular}{cc}
   $\mu$ & Landau point \n
   \sbentcore molecule composed of two parts: & \n
   $0.0$  & $107^{\circ}$ \n
   $1.2$  & $104^{\circ}$ \n
   $1.5$  & $103^{\circ}$ \n
   \sbentcore molecule composed of three parts: & \n
         & $\density$=0.026 $\density$=0.04 \n
   $0.0$ & $\dg{89}$ $\dg{89}$ \n
   $1.2$ & $\dg{86}$ $\dg{86}$  \n
   $1.6$ & $\dg{74}$ - $\dg{86}$ $\dg{74}$ - $\dg{86}$ \n
   $2.0$ & $\dg{63}$ - $\dg{86}$ $\dg{63}$ - $\dg{80}$ \n
   $2.8$ & $\dg{83}$ - $\dg{97}$ $\dg{82}$ - $\dg{92}$ \n
   \sbentcore molecule composed of two biaxial ellipsoids : & \n
   $0.0$  & $\dg{121}$ - $\dg{128}$ \n
  \end{tabular}
 \end{ruledtabular}
\end{table}
For the case of two-part molecule without the dipole moment Landau
point is found to be near $\gamma=107^{\circ}$, in agreement with
the hard-boomerang model \cite{teixeira}. The diagrams include two dipole strengths
$\mu=1.2$ and $\mu=1.5$ for which dipole-dipole interaction is
about $15\%$ and $22\%$, respectively, of the total potential
in ground state.
The bicritical point is shifted towards lower angles with increasing $\mu$,
which contrasts with the results of Monte Carlo simulation
for the  Lebwohl-Lasher lattice model\cite{mcbatesdipoles}, where
introduction of the dipoles resulted in a line of direct isotropic-biaxial
transitions.

For the non-polar three-part molecule, the Landau point is found to be
at $\gamma=89^{\circ}$ and is shifting to lower angles with increasing
dipole magnitude (Table~\ref{landauvsmu}) up to a point where the
dipole-dipole interactions make up $20\%$ ($\mu=1.4$) of the total potential in the ground state. Above that value the bicritical point changes into a line
of Landau points that widens with increasing $\mu$; for $\mu=1.6$ it covers the range of $\dg{12}$ and for
$\mu=2.0$ it extends for more than $\dg{20}$.

\begin{figure}
 \includegraphics{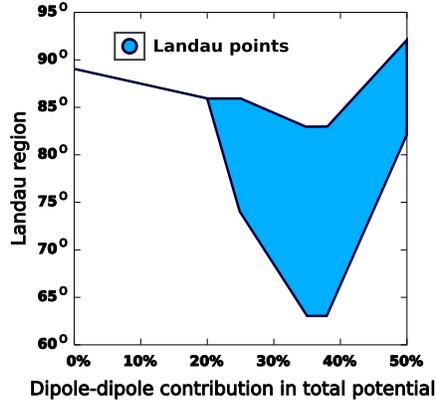}%
 \caption{
  \label{landauregions}
  The range of direct isotropic-biaxial transitions in bend angle, in function
  of dipole-dipole contribution in total potential.
 }
\end{figure}

The low $\gamma$ boundary practically does not change (for
lower density) and is equal to $\dg{86}$ for the dipole
strength $\mu \le 2.1$. Then the bicritical
region begins to shrink and is shifted towards higher angles. The highest
dipole studied was the one of $\mu=2.8$ for which $V_{DD}$ approaches
$50\%$ of the total potential energy. As can be seen from Fig.~\ref{diabramb2}
the bicritical line in that case is still getting shorter and moves towards higher bond angles. Fig.~\ref{landauregions} shows the evolution
of Landau region as function of the dipole magnitude, $\mu$.

The diagrams are presented for two densities such that the corresponding packing fraction is of the order of $0.3-0.4$. As can be seen from Table~\ref{landauvsmu} some differences appear with varying density for the strongest
dipoles ($\mu\geq 2.0$).
Namely the line of the direct isotropic-biaxial transitions shrinks for
higher density.

\begin{figure}
 \includegraphics{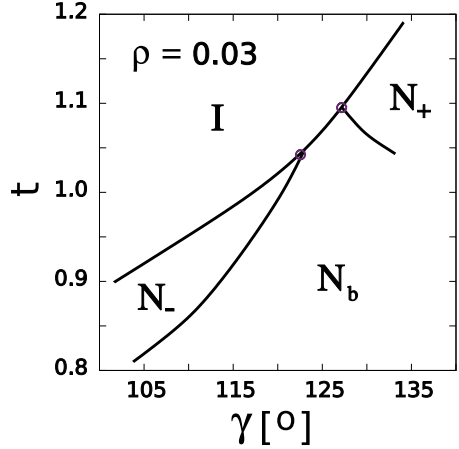}%
 \caption{
  \label{diagramb2bz2k}
  Bifurcation diagram for non-polar bent-core molecules modeled by two biaxial GB ellipsoids \cite{bfz}.
 }
\end{figure}

 Finally we take into account the model where the arms of the molecules
are biaxial. We are going to address the issue of observed disagreement between
the angles for which the Landau point appears experimentally ($\gamma=140^{\circ}$) and theoretically. Results presented below replace uniaxial GB arms  with their biaxial version developed
by Fava, Berardi and Zannoni \cite{bfz}. Interestingly, the Landau point in the biaxial model,
Fig.~\ref{diagramb2bz2k}, is replaced by a line of bicritical points even for the non-polar molecule. That line starts near $121^{\circ}$ and ends for $\gamma=128^{\circ}$.
The region becomes reduced to a single point with decreasing arm's biaxiality.

 Summarizing, we have presented a bifurcation study for a class of models with
characteristic features typical of the \sbentcore molecule. Using Density Functional Theory
we have retrieved the diagrams in low density
approximation. Analysis included two and three-part bend-cores with arms
modeled by GB interacting ellipsoids of uniaxial and biaxial symmetry.
The dipole-dipole interaction was added and the dipole strength influence
studied.
Non-polar uniaxial model revealed a single Landau point,
 in agreement with results for
hard molecules \cite{teixeira}. The deviation from uniaxial symmetry of the
arms resulted in transformation of the single bicritical point into a line of direct isotropic-biaxial
transitions.  The inclusion of the dipole-dipole interactions resulted in
shifting of the Landau point towards lower bond angles with increasing
dipole magnitude. For the case of
two-arm molecule stronger dipoles easier destabilized uniaxial nematic phase. For the three-part banana a line of the
bicritical points has emerged. The results suggest that there exists an optimal dipole range that makes the appearance of the biaxial phase most probable.

\begin{acknowledgments}
 Authors wish to thank Pawe\l{}   F. G\'ora and Micha\l{}  Cie\'sla for useful discussions. The work was supported by grant from MNiSW no N202 169 31/3455. The numerical analysis was performed using computer cluster at ICM under grant G27-8.
\end{acknowledgments}

 \bibliography{references}
\end{document}